\documentstyle[12pt,epsf]{article}

\font\tbf=cmbx10
\font\tenrm=cmr10
\font\twelverm=cmr12
\font\tit=cmti10
 1
 1
 1

\renewenvironment{thebibliography}[1]  
 { \tenrm
   \begin{list}{\arabic{enumi}.}
    {\usecounter{enumi} \setlength{\parsep}{0pt}
     \setlength{\itemsep}{0pt} 
     \setlength{\leftmargin}{20pt} 
     \sloppy
    }}{\end{list}}

\begin{document}
 {\bf 
   International Workshop on Electron Correlations and \\[1mm]
   Materials Properties,
   Crete, Greece, June 28 - July 3, 1998\\[1mm]
   (to be published in conference proceedings)}
 ~\\
 ~\\
 ~\\
 ~\\
{\bf DYNAMICAL ~ELECTRON ~CORRELATIONS ~IN ~METALS: \\ [1mm]
TB-LMTO ~AND ~MULTIBAND ~HUBBARD ~HAMILTONIAN}
 ~\\
 ~\\
 ~\\
\hspace*{25.4mm} V\'aclav Drchal, V\'aclav Jani\v{s}, and
                 Josef Kudrnovsk\'y\\[6pt]
\hspace*{25.4mm} Institute of Physics, Academy of Sciences of the Czech
Republic\\[4pt]   
\hspace*{25.4mm} Na Slovance 2, CZ-180 40 Praha 8, Czech Republic\\[4pt]  

 \begin{abstract}
   We study electronic properties of solids with correlated d electrons
   which could be described by a multiband Hubbard Hamiltonian in the
   weak-interaction case, $U/w<1$. The one-electron part of the many-body
   Hamiltonian is described by a tight-binding LMTO method. The many-body
   part is treated by non-selfconsistent FLEX-type approximations with
   adjusted chemical potential not to change the LMTO band filling. The
   calculated DOS gets narrower but is only little influenced by electron
   correlations at the Fermi energy.  A precursor of a satellite band in
   the paramagnetic bcc Ni is found at about 6 eV below the Fermi level in
   agreement with experiment.
 \end{abstract}
 ~\\
 ~\\
{\bf INTRODUCTION}\\

\twelverm

\hspace{9mm} 
A general scheme for determination of the electronic structure of solids
with correlated electrons is still missing.
The density functional formalism within the local density approximation
(LDA), in which the electron-correlation part is guessed on the basis of 
the homogeneous electron gas model, has proved to be a highly reliable 
method for the evaluation of the ground state properties \cite{HK}.
On the other hand, there are many examples of the failure of the LDA,
e.g., the study of the excitation spectra of solids or the evaluation
of the gap in insulators and semiconductors.
Another example of the failure of the LDA are solids whose electronic 
structure is better described in terms of 
atomic-like electronic states rather than in terms of the homogeneous 
electron gas model on which the LDA is based.

\hspace{9mm} 
The electronic properties of strongly correlated materials which cannot
be adequately described within the LDA are usually studied in the
framework of simplified models like, e.g., the single-band Hubbard model
(for a recent review see Ref.\cite{GKKR}).
The correlation effects in solids are usually classified in terms of
the ratio of the on-site Coulomb energy $U$ to the bandwidth $w$.
One distiguishes three regimes, namely, 
(i) the weak interaction case ($U/w < 1$, transition metals),
(ii) the intermediate interaction 
case ($U/w \approx 1$, metal-insulator transition regime and Kondo systems), and 
(iii) the strong interaction case ($U/w > 1$, rare-earth systems and wide gap 
solids). 
It should be mentioned that a general approach bridging all the above mentioned
cases is still missing even at the model level.

\hspace{9mm} 
Recently some progress has been made in the reformulation of the LDA
(the time-dependent LDA \cite{PGG}) which is useful in the study of 
excitation energies.
The so-called GW approximation (GWA) \cite{LH} was successfully 
applied to the gap problem in a number of semiconductors and insulators.
An alternative approach to the gap problem which is based on a
straightforward generalization of the LDA is the so-called
LDA+U method \cite{AZA}.
The GWA was also applied to metallic systems. 
Although computationally highly demanding, it yielded band narrowing 
in the photoemission spectra of nickel \cite{AG}, but it failed to
produce the well-known satellite structure below the main peak \cite{HWST}.
The presence of the satellite in Ni was succesfully explained in
the framework of the $T$-matrix formalism developed recently on
{\it ab initio} level (see Ref.~\cite{SAK} and references therein
for previous semiempirical approaches to this problem).
An alternative  approach to the satellite and the band narrowing in nickel 
is based on a three-body scattering approximation that employs the
Faddeev equations \cite{FA1,FA2}.

\hspace{9mm} 
In a recent paper \cite{LK} the so-called LDA++ method was introduced
which extends the LDA+U method by accounting for dynamical electron
correlations.
The starting point of the LDA++ method is a multiband Hubbard 
Hamiltonian whose one-electron part is identified with the LDA
Hamiltonian with a subtracted double counting correction for the
average Coulomb interactions in the LDA.
The LDA++ approximates solutions in different correlation limits
by different many-body approximations, namely it employs the so-called
Hubbard I solution \cite{HI} in the strong-interaction limit, the 
iterated perturbation theory (IPT) within the dynamical mean-field 
theory (DMFT) \cite{GKKR} for the intermediate-interaction case,
and the so-called fluctuation exchange approximation (FLEX) 
\cite{BS} in the weak-interaction limit.

\hspace{9mm} 
In the present paper we follow the basic approach of the LDA++
but with some conceptual as well as computational differences.
In particular, the main effort is directed towards a formulation
of the problem in a way suitable for future generalizations 
to random alloys and their surfaces and interfaces.
This means that the formalism is strictly based on Green functions
which allow for configurational averaging in the case of alloys.
The starting point is a multiband Hubbard Hamiltonian (MBHH) whose 
parameters are naturally determined from the corresponding tight-binding
linear muffin-tin orbital (TB-LMTO) Hamiltonian (for a review
of the TB-LMTO method and its applications to random surfaces and interfaces
see a recent book, Ref.~\cite{TDKSW}).
The basic approximation adopted here is the assumption on the local 
(site-diagonal, or wave-vector independent) selfenergy which is
reasonably well justified for transition metals and their alloys
\cite{SAS}.
The local approximation for the selfenergy is also required from the 
formal point of view for the extension of the theory to the case of 
random alloys within the coherent potential approximation 
(CPA) \cite{VKE}.
We employ the single-channel approximations (FLEX) of the canonical 
perturbation theory to solve the many-body part of the problem.
 ~\\
 ~\\
 ~\\
{\bf THEORY}\\
 
\hspace{9mm}
The electronic structure determined within the LDA is described by a
TB-LMTO Hamiltonian with the overlap matrix equal to unity
\begin{equation}
H^{\rm LMTO} = \sum_{{\bf R}\lambda,{\bf R'}\lambda'} \,
|{\bf R}\lambda\rangle \, H^{\rm LMTO}_{{\bf R}\lambda,{\bf R'}\lambda'}
\, \langle {\bf R'}\lambda'| \, , 
\label{hlmto1}
\end{equation}
where
\begin{equation}
H^{\rm LMTO}_{{\bf R}\lambda,{\bf R'}\lambda'} =
C_{{\bf R}\lambda\lambda'} \, \delta_{\bf RR'} +
\Delta^{1/2}_{{\bf R}\lambda} \, 
S_{{\bf R}\lambda,{\bf R'}\lambda'}^{\gamma}
\, \Delta^{1/2}_{{\bf R'}\lambda'} \, .
\label{hlmto2}
\end{equation}
Here, ${\bf R}$ is the site index, $\lambda=(L\sigma) = (\ell m \sigma)$
is the spinorbital index,
$L=(\ell m)$ is the orbital index, $\sigma$ is the $z$-projection 
of the spin, $C$, $\Delta$, and $\gamma$ are site-diagonal matrices of 
potential parameters and $S^{\gamma}$ is the matrix of structure 
constants in the orthogonal LMTO representation
\begin{equation}
S_{{\bf R}\lambda,{\bf R'}\lambda'}^{\gamma}=
[S^0(1-\gamma S^0)^{-1}]_{{\bf R}\lambda,{\bf R'}\lambda'}=
\left[S^{\beta}\left(1-(\gamma-\beta) 
S^{\beta}\right)^{-1}\right]_{{\bf R}\lambda,{\bf R'}\lambda'}.
\label{ortsc}
\end{equation}
Here $S^0$ is a matrix of canonical structure constants,
$S^{\beta}$ is a matrix of screened structure constants,
and $\beta$ is a site-diagonal matrix of screening constants.

\hspace{9mm} 
The parameters of a multiband Hubbard Hamiltonian
in second quantization with creation $(a^{+}_{{\bf R}\lambda})$
and destruction $a_{{\bf R}\lambda}$ operators
\begin{equation}
H^{\rm Hubb} = \sum_{{\bf R}\lambda,{\bf R'}\lambda'} 
t_{{\bf R}\lambda,{\bf R'}\lambda'} \,
a^{+}_{{\bf R}\lambda} \, a_{{\bf R'}\lambda'} +
\sum_{{\bf R},\lambda,\lambda'} U_{{\bf R}\lambda\lambda'} \,
n_{{\bf R}\lambda} \, n_{{\bf R}\lambda'}
\label{mbhh}
\end{equation}
are found from the LDA calculations.
The density operator $n_{{\bf R}\lambda}=a^{+}_{{\bf R}\lambda} 
\, a_{{\bf R}\lambda}$.
The hopping integrals 
\begin{equation}
t_{{\bf R}\lambda,{\bf R'}\lambda'} =
\Delta^{1/2}_{{\bf R}\lambda}\, S_{{\bf R}\lambda,{\bf R'}\lambda'}^{\gamma}
\, \Delta^{1/2}_{{\bf R'}\lambda'} \, , \quad {\bf R} \neq {\bf R'}
\label{hopp}
\end{equation}
are identified with the site off-diagonal elements of (\ref{hlmto1}),
while the atomic levels $t_{{\bf R}\lambda,{\bf R}\lambda'} = 
\epsilon_{{\bf R}\lambda\lambda'} = \epsilon_{{\bf R}\lambda}
\delta_{\lambda,\lambda'}$
are determined from the condition that the average occupancy 
$\bar{n}_{{\bf R} \lambda}$ of the state $|{\bf R}\lambda\rangle$ 
as calculated within the LDA and within the Hartree-Fock
approximation for multiband Hubbard Hamiltonian (\ref{mbhh}) are identical. 
It means
\begin{equation}
\epsilon_{{\bf R}\lambda} = C_{{\bf R}\lambda} - \Sigma_{{\bf R}\lambda}^{\rm HFA}
= C_{{\bf R}\lambda} - 
\sum_{\lambda'} U_{{\bf R}\lambda\lambda'} \, \bar{n}_{{\bf R}\lambda'} \, .
\label{hfalda}
\end{equation}

The Hubbard interaction parameter $U_{{\bf R}\lambda\lambda'}$ is determined 
within the LDA usually as a second derivative of $E_{tot}$ with 
respect to occupancies of the interacting states $\lambda,\lambda'$.
Due to the Pauli principle, $U_{{\bf R}\lambda\lambda'}=
U_{{\bf R}\lambda\lambda'}(1-\delta_{\lambda\lambda'})$.
Here we consider translationally invariant solids with one atom per unit cell.
The hopping integrals then depend only on the difference ${\bf R-R'}$.
The on-site levels $\epsilon_{{\bf R}\lambda}$ as well as the pair interactions
$U_{{\bf R}\lambda\lambda'}$ are independent of ${\bf R}$.

\hspace{9mm} 
We study four different approximations of the canonical many-body 
perturbation theory, namely, (i) the second order perturbation theory (SOPT), and 
three approximations of infinite order that correspond to electron-electron (hole) 
scatterings in a single channel, namely, (ii) the T-matrix approximation (TMA)
which describes electron-electron scatterings, (iii) the random-phase 
approximation (RPA) corresponding to electron-hole scatterings, and 
(iv) the GW-approximation (GWA) in which the bare interaction $U$ is replaced
by a (complex and energy-dependent) interaction $W(E)$ (double wavy line) renormalized 
by repeated excitations of electron-hole pairs (polarization bubbles).
The graphs for one-particle selfenergy are shown in Fig.~1.
We assume 

\vspace*{-18mm}
\epsfxsize=187mm
\vglue 5mm
\epsffile{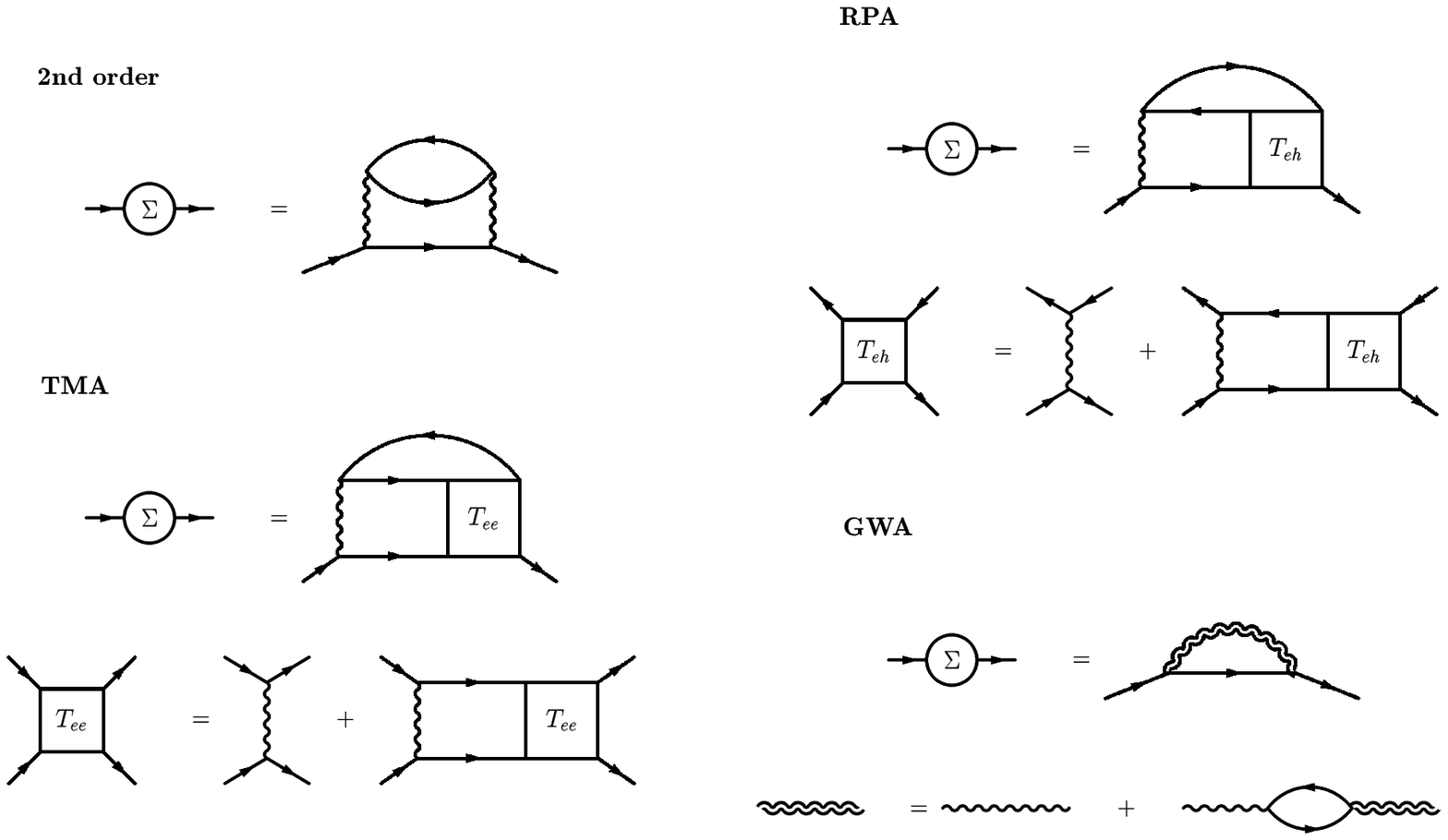}
\vspace*{-155mm}
{\tbf Figure 1.}
{\tenrm Graphs for one-particle selfenergy in second order perturbation
theory (SOPT), T-matrix approximation (TMA), random-phase approximation (RPA),
and GW-approximation (GWA).}
\vspace*{2mm}

\twelverm
a local approximation, i.e., the selfenergy is independent
of the {\bf k}-vector, which means that it is diagonal in the site 
representation, $\Sigma_{{\bf R}\lambda,{\bf R'}\lambda'}=
\Sigma_{{\bf R},\lambda\lambda'} \delta_{{\bf R},{\bf R'}}$.
We use the technique of causal Green functions at $T=0$ in our calculations.
It follows from the adopted construction of the on-site levels in (\ref{hfalda})
that the Hartree-Fock part of the one-electron selfenergy is already contained 
in the atomic levels $\epsilon_{{\bf R}\lambda}$.
Consequently, the expansions of the selfenergy start from the terms quadratic
in the Hubbard $U$. 
The unperturbed one-particle Green function corresponding to the Hartree-Fock 
approximation (HFA) applied to $H^{\rm Hubb}$ is identical with the Green 
function of $H^{\rm LMTO}$.

\hspace{9mm} 
We assume a cubic symmetry of the lattice and confine ourselves to $s$, $p$,
and $d$ states.
In addition, we neglect the pair correlations in $s$ and $p$ states and for
simplicity we assume $U_{\lambda\lambda'}=U(1-\delta_{\lambda\lambda'})
\delta_{\ell,2}$, i.e., we neglect exchange interactions and consider only 
the pair interactions between the $d$-states.
The Green functions and selfenergy are then diagonal in the spinorbital 
indices $\lambda$.
The selfenergy can acquire only two values that correspond to $e_g$ and $t_{2g}$ 
representations of the cubic point group.

\hspace{9mm} 
As an example we give here the derivation of basic equations for the TMA
which is an extension of the derivation given in Ref.~\cite{VD79}
to the multiband case.
The basic equations of the TMA in the language of causal quantities 
(superscript $c$) read
\begin{equation}
i\Psi^{(c)}_{\lambda\lambda'}(E)=
-\int^{\infty}_{-\infty} \frac{{\rm d}\omega}{2\pi} \,
G^{(c)}_{\lambda}(E-\omega)\, G^{(c)}_{\lambda'}(\omega) \, ,
\label{ceegf}
\end{equation}
\begin{equation}
T^{(c)}_{\lambda\lambda'}(E)=
\frac{U}{1-U\Psi^{(c)}_{\lambda\lambda'}(E)}-U \, ,
\label{ctm}
\end{equation}
\begin{equation}
-i\Sigma^{(c)}_{\lambda}(E)=
\sum_{\lambda'} \, (-1) \int^{\infty}_{-\infty}
\frac{{\rm d}\omega}{2\pi} \,
T^{(c)}_{\lambda\lambda'}(E+\omega) \, G^{(c)}_{\lambda'}(\omega) \, ,
\label{ctmase}
\end{equation}
where $\Psi^{(c)}_{\lambda\lambda'}(E)$ is the two-particle propagator,
$T^{(c)}_{\lambda\lambda'}(E)$ is the electron-electron T-matrix, and
$\Sigma^{(c)}_{\lambda}(E)$ is the one-electron selfenergy.
The factor $(-1)$ in (\ref{ctmase}) is due to a closed fermion loop.
In numerical calculations, it is advantageous to replace the causal 
quantities by the retarded ones (without superscript) as follows
\begin{equation}
\Psi^{(c)}_{\lambda\lambda'}(E)=
{\rm Re}\Psi_{\lambda\lambda'}(E) + i \,{\rm sgn}(E-2\mu)\,
{\rm Im}\Psi_{\lambda\lambda'}(E) \, ,
\label{creegf}
\end{equation}
\begin{equation}
T^{(c)}_{\lambda\lambda'}(E)=
{\rm Re}T_{\lambda\lambda'}(E) + i\, {\rm sgn}(E-2\mu)\,
{\rm Im}T_{\lambda\lambda'}(E) \, ,
\label{crtm}
\end{equation}
\begin{equation}
\Sigma^{(c)}_{\lambda}(E)=
{\rm Re}\Sigma_{\lambda}(E) + i\, {\rm sgn}(E-\mu)\,
{\rm Im}\Sigma_{\lambda}(E) \, ,
\label{crtmase}
\end{equation}
\begin{equation}
G^{(c)}_{\lambda}(E)=
{\rm Re}G_{\lambda}(E) + i\, {\rm sgn}(E-\mu)\,
{\rm Im}G_{\lambda}(E) \, .
\label{crgf}
\end{equation}
By inserting (\ref{creegf})-(\ref{crgf}) into (\ref{ceegf})-(\ref{ctmase})
we find equations for the retarded quantities which are holomorphic 
in the upper halfplane of the complex energy $z$ and vanish for $z \rightarrow 0$.
It is therefore sufficient to calculate only their imaginary parts because
the real parts can be found using a dispersion relation
\begin{equation}
X(z)= \int^{\infty}_{-\infty} \frac{{\rm d}\omega}{\pi} \,
\frac{{\rm Im} X(\omega)} {\omega-z} \, , \quad
\label{hilbtr}
\end{equation}
which allows to determine the real and imaginary parts of $X(z)$ 
for complex $z$ as well as for real values of the energy.
The equations for the retarded quantities read
\begin{equation}
{\rm Im}\Psi_{\lambda\lambda'}(E)=
-{\rm sgn}(E-2\mu) \, 
\int^{\mu}_{E-\mu} \frac{{\rm d}\omega}{\pi} \,
{\rm Im} G_{\lambda}(E-\omega)\, {\rm Im} G_{\lambda'}(\omega) \, ,
\label{reegf}
\end{equation}
\begin{equation}
T_{\lambda\lambda'}(E)=
\frac{U}{1-U\Psi_{\lambda\lambda'}(E)}-U \, ,
\label{rtm}
\end{equation}
\begin{equation}
{\rm Im}\Sigma_{\lambda}(E)=
-\sum_{\lambda'} \,  \int^{\mu}_{2\mu -E}
\frac{{\rm d}\omega}{\pi} \,
{\rm Im}T_{\lambda\lambda'}(E+\omega) \, 
{\rm Im}G_{\lambda'}(\omega) \, .
\label{rtmase}
\end{equation}
In the derivation of (\ref{reegf})-(\ref{rtmase}) we have used analytic 
properties of the retarded quantities from which follow useful relations 
between convolutions of real and imaginary parts, as, for example
\begin{equation}
\int^{\infty}_{-\infty} \frac{{\rm d}\omega}{2\pi} \,
{\rm Re} G_{\lambda}(E-\omega)\, {\rm Re} G_{\lambda'}(\omega) =  
-\int^{\infty}_{-\infty} \frac{{\rm d}\omega}{2\pi} \,
{\rm Im} G_{\lambda}(E-\omega)\, {\rm Im} G_{\lambda'}(\omega) \, .
\label{idegf}
\end{equation}

\hspace{9mm} 
The equations for the other approximations can be derived in a similar way.
An expression for the renormalized pair interaction $W(E)$ requires matrix
inversion even in the simplified case 
$U_{\lambda\lambda'}=U(1-\delta_{\lambda\lambda'})$ assumed here, since
\begin{equation}
\sum_{\lambda''} \, [\delta_{\lambda,\lambda''}-U_{\lambda\lambda''}
\Phi_{\lambda''}(E)] \, W_{\lambda''\lambda'}(E)=U_{\lambda\lambda'}
\, ,
\label{wavylin}
\end{equation}
where $\Phi_{\lambda}(E)$ is the electron-hole bubble and
\begin{equation}
{\rm Im}\Phi_{\lambda}(E)=
-\int^{\mu}_{\mu-E} \frac{{\rm d}\omega}{\pi} \,
{\rm Im} G_{\lambda}(E+\omega)\, {\rm Im} G_{\lambda}(\omega) \, .
\label{rbubble}
\end{equation}

\hspace{9mm} 
The retarded one-electron Green function of the interacting system
is given by the resolvent
\begin{equation}
G(z)=[z-H^{\rm eff}(z)]^{-1}
\label{gfint}
\end{equation}
of an effective one-electron Hamiltonian with the matrix elements
\begin{equation}
H^{\rm eff}_{{\bf R}\lambda,{\bf R'}\lambda'}(z) =
H^{\rm LMTO}_{{\bf R}\lambda,{\bf R'}\lambda'} +
\Sigma_{\lambda}(z) \, \delta_{\bf RR'} \, \delta_{\lambda\lambda'} \, .
\label{heff}
\end{equation}
The density of states (DOS) is defined as
\begin{equation}
\rho(E)=-\frac{1}{\pi} \, {\rm Im} \sum_{\lambda} \, 
G_{{\bf R}\lambda,{\bf R}\lambda}(E+i0) \, .
\label{dos}
\end{equation}
The set of equations has to be completed by an equation for the chemical
potential $\mu$
\begin{equation}
\int^{\mu}_{-\infty}{\rm d}E \, \rho(E) = n \, ,
\label{chempot}
\end{equation}
where $n$ is the number of electrons per one site.

\hspace{9mm} 
The spectral density
\begin{equation}
A({\bf k},E)=-\frac{1}{\pi} {\rm Im} \sum_{\lambda} \, 
G_{\lambda\lambda}({\bf k},E+i0)
\label{ake}
\end{equation}
is expressed in terms of the lattice Fourier transform of $G_{\bf RR'}(z)$.
 ~\\
 ~\\
 ~\\
{\bf NUMERICAL IMPLEMENTATION}\\

\hspace{9mm}
The formalism developed in the previous section was applied to the 
case of the paramagnetic fcc-Ni and the paramagnetic bcc-Fe which 
belong to weakly interacting systems.
There is no exact way of determining the Hubbard interaction 
parameter $U$ for the interaction among the $d$-electrons.
One can find a broad variety of values in the literature.  
In the present calculations we have adopted $U=0.18$~Ry for
the paramagnetic fcc-Ni which emphasizes its atomic-like character,
while the value $U=0.1$~Ry used for the paramagnetic bcc-Fe
is close to the value derived from experiment (see Ref. \cite{SAS}).
The ratio $U/w$ is larger for Ni as compared to Fe because the Ni 
bandwidth is smaller and the corresponding $U$ is also larger.
Therefore many-body effects will be more pronounced in Ni metal as
compared to Fe. 
The smaller value of $U/w$ in iron is, however, partly offset 
by a larger number of holes so that one can still expect non-negligible
influence of many-body effects on one-particle states.
The parameters of the one-particle part of the MBHH were obtained from
the TB-LMTO assuming experimental lattice constants and the
Ceperley-Alder form of the exchange-correlation potential.
We note that the TB-LMTO method employs the so-called atomic sphere
approximation (ASA) which gives an accurate description of ground state
properties of transition metals, their alloys and surfaces.
For more details we refer to a recent book, Ref.~\cite{TDKSW}.

\hspace{9mm}
The elements of the site-diagonal Green function corresponding to
the one-particle part of the MBHH, $G_{\bf RR}(E)$, needed for the 
many-body calculations, are obtained from the site-diagonal elements 
of the auxiliary Green function, $g_{\bf R R}(z), z=E+i0$, as
\begin{eqnarray} \label{pgf}
G_{{\bf R}\lambda,{\bf R}\lambda'}(z) &=& 
[(z - H^{\rm LMTO})^{-1}]_{{\bf R}\lambda,{\bf R}\lambda'}
\nonumber \\
&=&\lambda_{{\bf R}\lambda}(z) \; \delta_{\lambda\lambda'} + 
\mu_{{\bf R}\lambda}(z) \, g_{{\bf R}\lambda,{\bf R}\lambda'}(z) \, 
\mu_{{\bf R}\lambda'}(z) \, ,
\end {eqnarray}
where
\begin{equation} \label{agf}
g_{{\bf R}\lambda,{\bf R}\lambda'}(z) = 
\frac{1}{N} \sum_{\bf k} \, [({\cal{P}}(z) -
S({\bf k)})^{-1}]_{\lambda\lambda'} \, .
\end {equation}
In (\ref{agf}), the sum runs over the Brillouin zone, $N$ denotes 
the number of lattice sites, and the site-diagonal quantities 
$\lambda_{{\bf R}\lambda}(z)$ and $\mu_{{\bf R}\lambda}(z)$ 
are functions of the potential parameters $C_{\bf R}$,
$\Delta _{\bf R}$, and $\gamma_{\bf R}$ (see Ref.~\cite{TDKSW}
for more details).
Finally, ${\cal P}_{{\bf R}\lambda}(z)$ is the potential function
which is expressed in terms of potential parameters as
\begin{equation} \label{pf0}
{\cal P}_{{\bf R}\lambda}(z) = \frac{z - C_{{\bf R}\lambda}}
{\Delta_{{\bf R}\lambda} + (\gamma_{{\bf R}\lambda} - 
\beta_{{\bf R}\lambda})(z - C_{{\bf R}\lambda})} \, .
\end{equation}
The calculations are performed first along a line in the complex energy 
plane ($z=E+i \epsilon$, $|\epsilon|=0.01$~Ry), and the results are analytically 
deconvoluted back to the real axis \cite{HVE}.
In this way, the poles present in Eq.~(\ref{agf}) are avoided.
Once the imaginary part of $G_{{\bf R}\lambda,{\bf R}\lambda}(E+i0)$ 
is known, the selfenergy is determined from Eqs.~(\ref{reegf})-(\ref{rtmase}).
This procedure is non-selfconsistent with respect to the Green function 
and the selfenergy.
However, we determine the Fermi energy from (\ref{hfalda}) in a 
selfconsistent manner in order that the number of particles is conserved.
In the last step we evaluate the quantities of interest, in particular,
the density of states $\rho(E)$, Eq.~(\ref{dos}), and the spectral density
$A({\bf k},E)$, Eq.~(\ref{ake}).
To this end we need to perform calculations indicated above but now for
the Hamiltonian $H^{\rm eff}$, Eq.~(\ref{heff}), rather then for $H^{\rm LMTO}$.
In other words, the potential function (\ref{pf0}) is substituted by a
function of the same form in which $C_{{\bf R}\lambda}$ is replaced by
\begin{equation} \label{effC}
\widetilde{C}_{{\bf R}\lambda}(z)=C_{{\bf R}L} + 
\Sigma_{{\bf R}\lambda}(z) \, .
\end{equation}
The calculations are again performed along the line $z=E+i \epsilon$ in 
the complex energy plane and then deconvoluted to the real axis.
The selfenergy in the complex energy plane is obtained 
from its imaginary part on the real axis determined above using 
dispersion relation (\ref{hilbtr}).

\hspace{9mm} 
We have used 2480 and 1360 ${\bf k}$-points in the irreducible Brillouin 
zone to perform the integrations in Eq.~(\ref{agf}) for the fcc and bcc
cases, respectively.
We note that such a high number of ${\bf k}$-points is only needed to 
reduce the oscillations in the DOS tails which are due to the discretization 
of $\bf k$-space, while for the determination of the selfenergy
it is sufficient to use much smaller number of ${\bf k}$-points.
We have verified that the Fermi energy and the selfenergy obtained using
280 and 240 ${\bf k}$-points give essentially the same result.
The step in the energy was 0.005~Ry.
A typical run for 750 energy points and 2500 $\bf k$-points requires
about 5 minutes on a medium workstation and the selfconsistency with
respect to the Fermi energy is achieved typically after 5 iterations
in the weak-interaction case.
~\\
~\\
~\\
{\bf RESULTS AND DISCUSSION}\\

\hspace{9mm} 
In Fig.~2 we compare the DOS of the paramagnetic fcc-Ni evaluated in 
the non-selfconsistent SOPT and in the non-selfconsistent TMA.
In each case the results are compared with the corresponding LDA-DOSs.
The following points are to be mentioned: 
(i) we observe a narrowing of the one-particle (LDA) bands, which is 
in agreement with experimental data \cite{HWST}; 
(ii) we note sharp structures close to the Fermi energy which are
only weakly influenced by electron interactions. 
The DOS at the Fermi energy is thus rather similar to that obtained 
in the LDA. 
Note that a large value of the DOS at the Fermi level is a precursor 
of a magnetic state at $T=0$~K (the Stoner criterion) which does 
not seem to be influenced by electron interactions;

\epsfxsize=120mm
\vglue 1mm
\epsffile{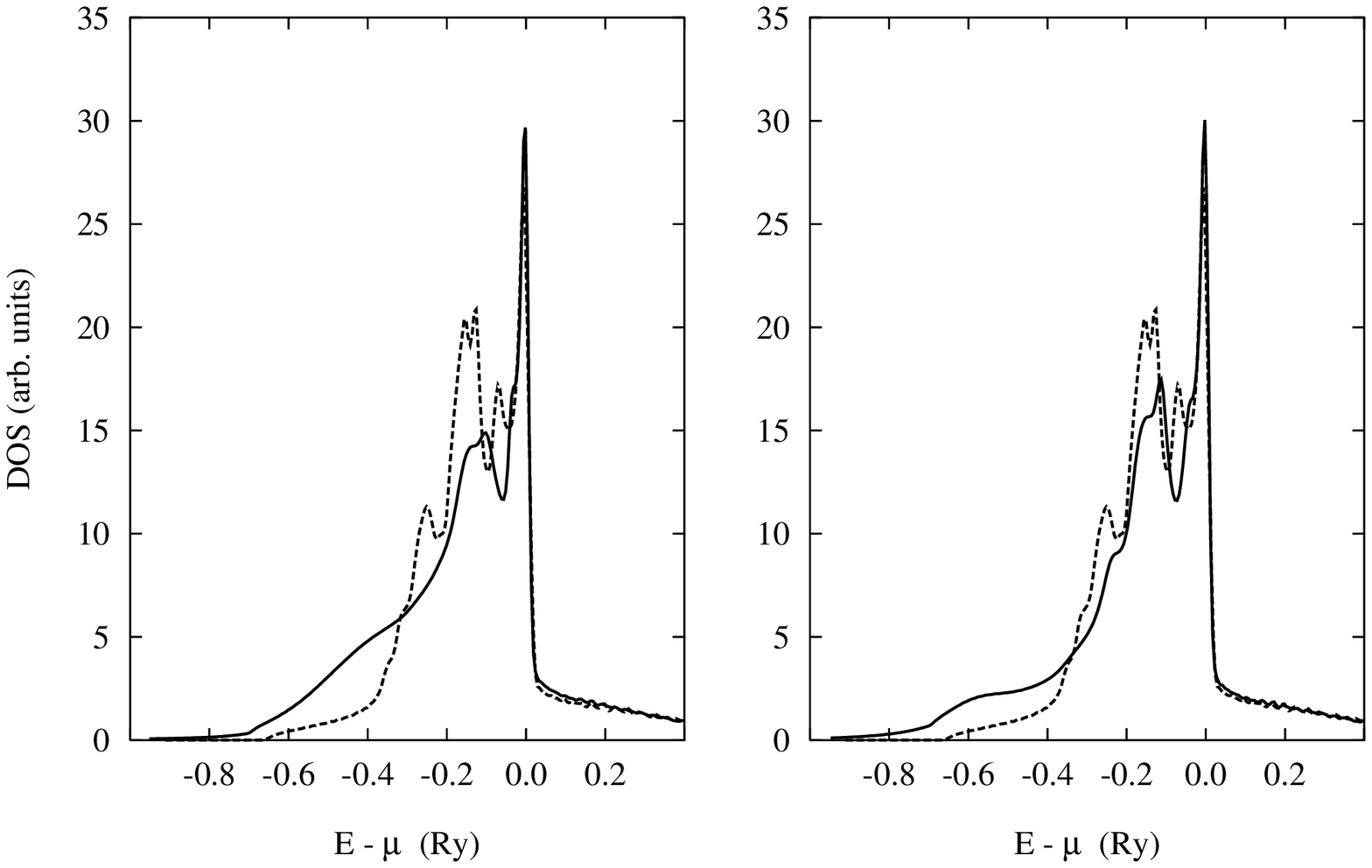}
\vspace*{-15mm}
{\tbf Figure 2.}
{\tenrm Densites of states for a paramagnetic fcc-Ni within the LDA 
        (dashed lines) 
       and non-selfconsistent 2nd-order perturbation theory (left frame, full 
       line) and non-selfconsistent T-matrix approximation (right frame, full
       line) for the Hubbard parameter $U=0.18$ Ry. }

\hspace{9mm} 

(iii) the $3d-3d$ electron interactions in a partly filled band also give 
rise to a broad dispersionless satellite which appears below the Fermi 
energy. 
This satellite is positioned at about 6 eV below the Fermi energy
in the SOPT in agreement with experiment \cite{HWST}. 
It should be noted, however, that its position depends on the value 
of the interaction parameter $U$ used in the calculations. 
The results for the SOPT are in a reasonable agreement 
with similar calculations of Steiner et al. \cite{SAS};
(iv) the effect of electron interactions for the same value of the 
Hubbard parameter $U$ is stronger in the TMA. 
In particular, the satellite structure is more pronounced and shifted 
deeper below the Fermi energy. 
This is due to the denominator of the T-matrix (see Eq.~\ref{rtm}) 
which is missing in the SOPT expression.

\hspace{9mm} 
We have performed similar calculations also for the RPA and the GWA.
The results for the RPA are very similar to the results of the SOPT.
The reason is that the $k$-order diagram of the RPA is of order of 
$O(n^k_{\rm h})$, where $n_{\rm h}$ is the number of holes per site
which is small for Ni so that the first term $(k=2)$ in the 
perturbation series, which is identical to the SOPT diagram, dominates.
On the other hand, the effects of electron interactions (band narrowing 
and the shift of the satellite below the Fermi energy) are strongest 
within the GWA. 

\hspace{9mm} 
A deeper insight into the nature of the electron states influenced by 
electron-electron interactions could be obtained from spectral densities,
or ${\bf k}$-resolved densities of states.
We note that the DOS and the spectral densities are, with exception
of the transition matrix elements, proportional to the angle-integrated
and angle-resolved photoemission spectra, respectively.
Despite the fact that the effect of the transition matrix elements may
be important, such quantities are still very helpful in understanding
the influence of electron interactions on the electron states.
In Fig.~3 we present the spectral densities of Ni metal 
calculated within the SOPT for $\bf k$=(0,0,0) ($\Gamma$-point) 
and for ${\bf k}=\frac{2\pi}{a}(1,0,0)$ (X-point) 
in the BZ of the fcc-lattice together with the LDA spectral 

\epsfxsize=120mm
\epsffile{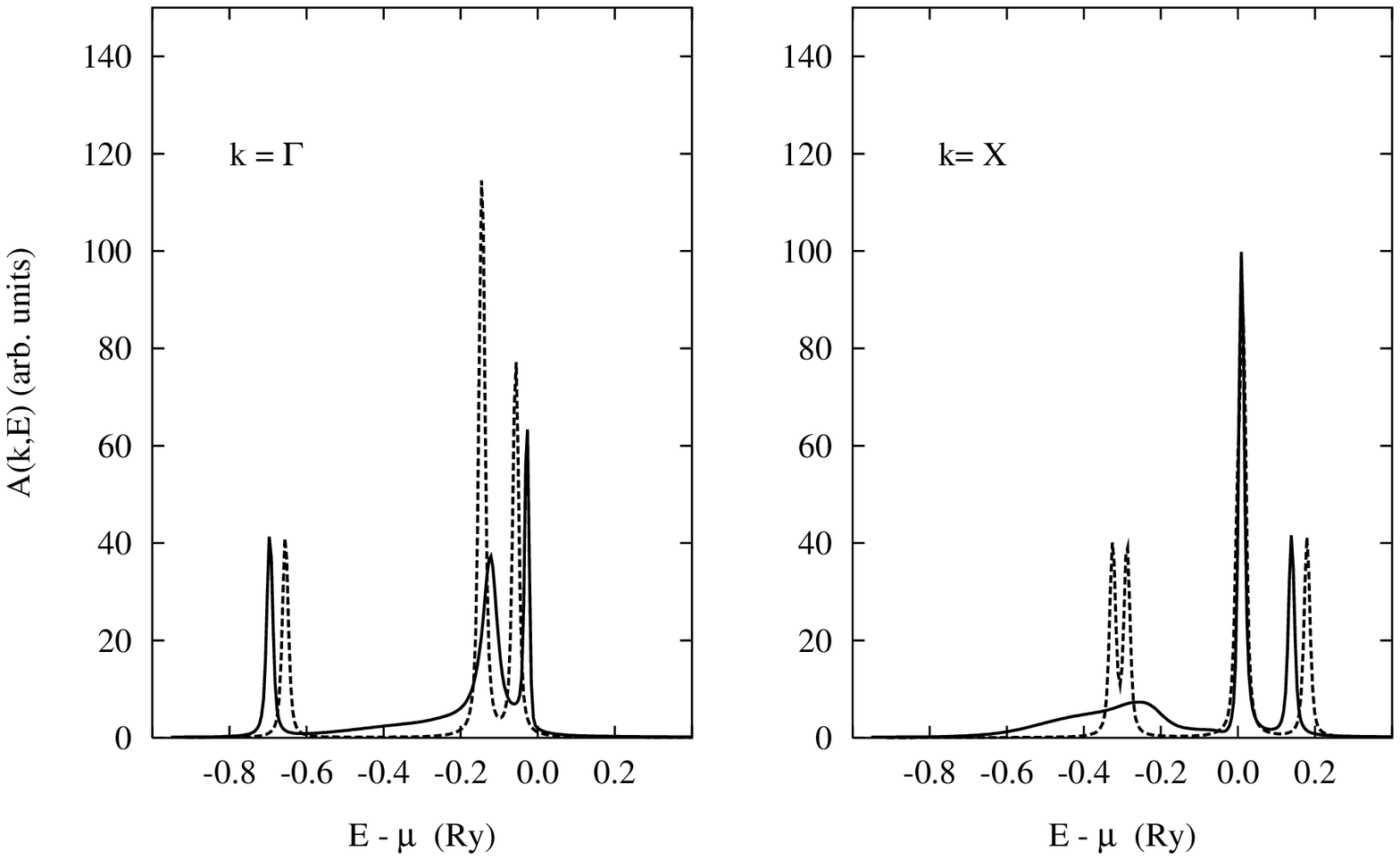}
\vspace*{-15mm}
{\tbf Figure 3.}
{\tenrm Spectral densities for a paramagnetic fcc-Ni within the LDA 
       (dashed lines)
       and non-selfconsistent 2nd-order perturbation theory (full lines) at 
       ${\bf k}=\Gamma$ (left frame) and at ${\bf k}$=X (right frame) for 
       Hubbard parameter $U=0.18$ Ry.}

\hspace{9mm} 

densities.
In calculating the
spectral densities we added a small imaginary part 
(0.005 Ry) to the energy to avoid too sharp peaks, in particular for
the LDA curves with no damping due to the selfenergy.
Three LDA peaks for ${\bf k}=\Gamma$ correspond to three bands, one
of the $s$-character at $E\approx -0.75$~Ry and two other corresponding
to the $d$-states around $E\approx -0.2$~Ry.
At the X-point we have in the LDA two peaks around $E\approx -0.3$~Ry 
corresponding to the $d$-states of which the lower one has a strong admixture 
of the $s$-states, the other two $d$-bands at the Fermi energy (not resolved 
here because of their closeness and the finite imaginary part of the energy 
used in the calculations), and one $sp$-band above the Fermi energy (see
Fig.~4). 
The influence of electron interactions results in a shift and broadening
of the electron states, but the effect is anisotropic with respect to
${\bf k}$-vector and it depends also strongly on the energy region.

\epsfxsize=120mm
\vglue 6mm
\epsffile{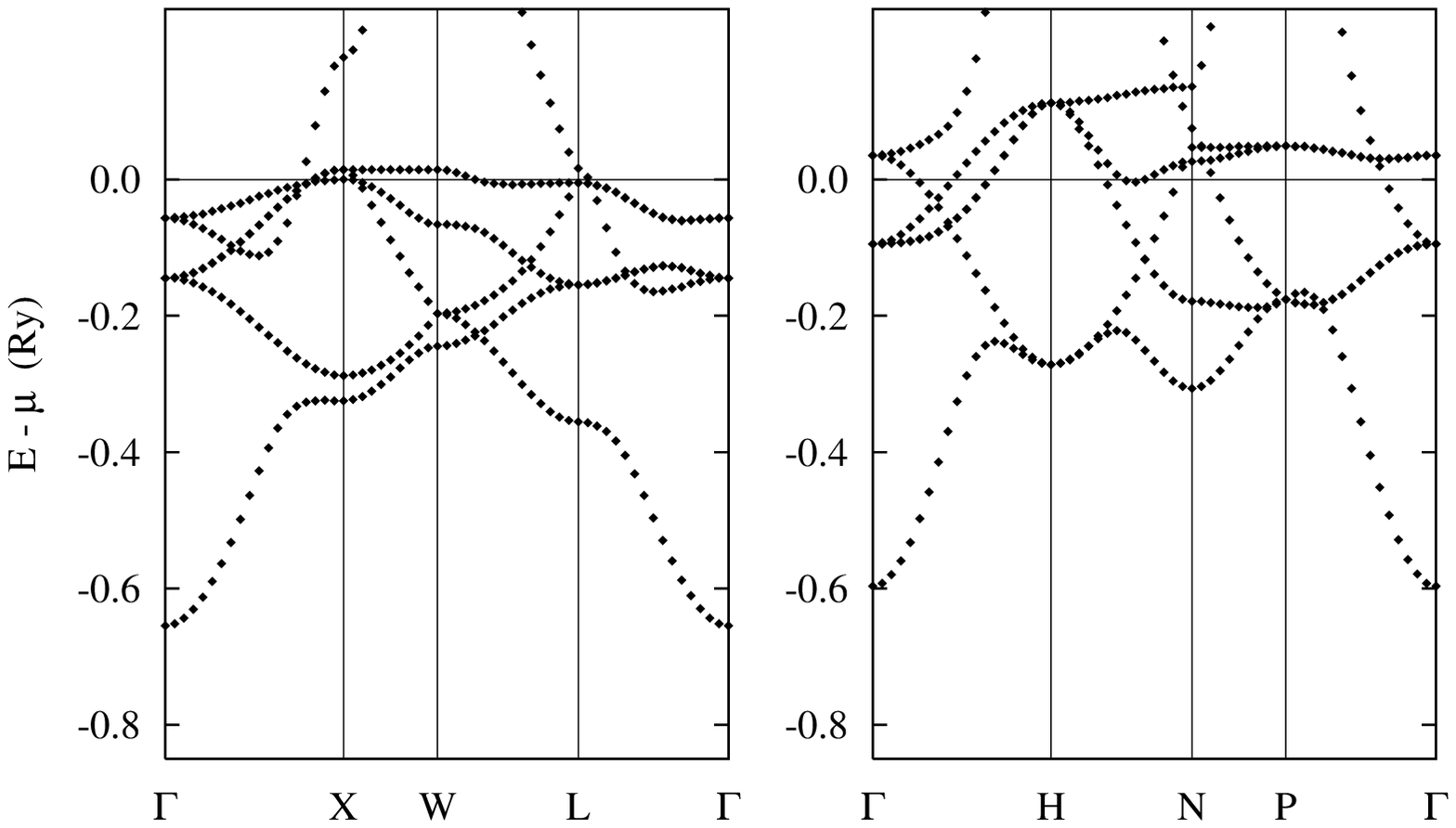}
\vspace*{-15mm}
{\tbf Figure 4.}
{\tenrm The LDA-bandstructures of paramagnetic fcc-Ni (left frame) and
        of paramagnetic bcc-Fe (right frame) along the lines of high
        symmetry in reciprocal space.}

\hspace{9mm} 
\epsfxsize=140mm
\vglue 1mm
\epsffile{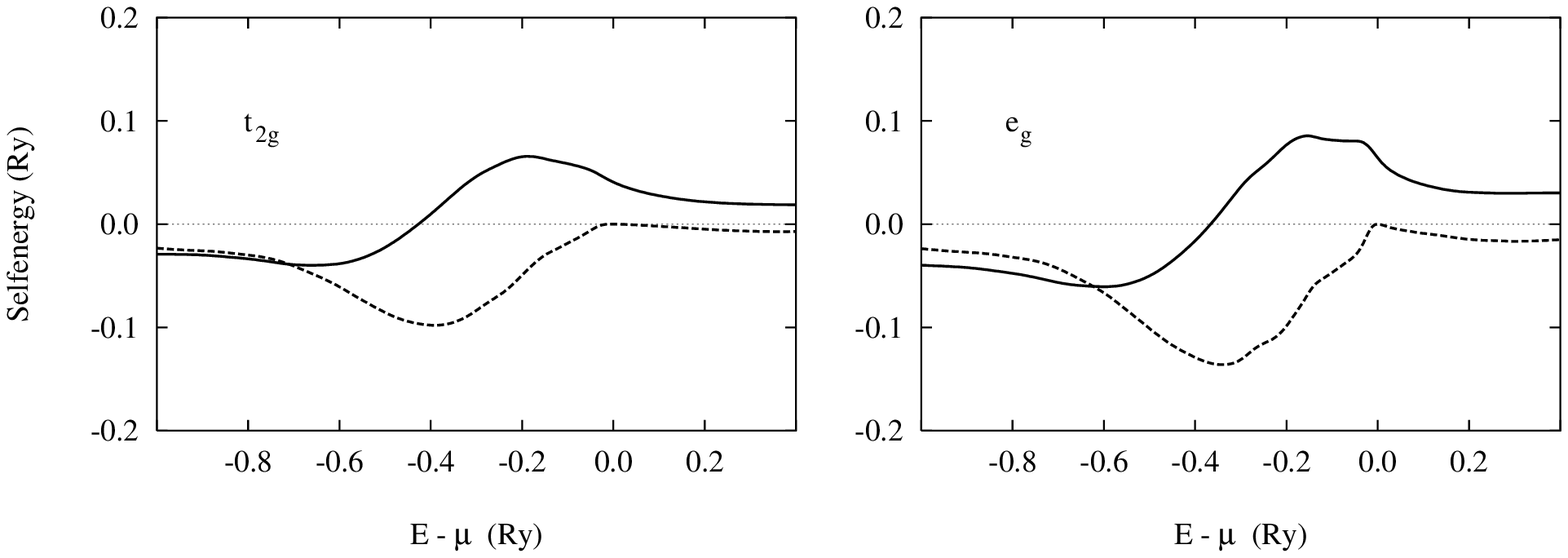}
\vspace*{-15mm}
{\tbf Figure 5.}
{\tenrm  Real (full lines) and imaginary (dashed lines) parts of the 
         selfenergy for a paramagnetic fcc-Ni for the non-selfconsistent 
         2nd-order perturbation theory: $t_{2g}$-symmetry (left frame) and 
         $e_{g}$-symmetry (right frame) for Hubbard parameter $U=0.18$ Ry.}

\hspace{9mm} 

To understand such a behavior in more detail, we plot in Fig.~5 the real
and imaginary parts of the selfenergy $\Sigma_{\lambda}(E)$.
Note that our approximations obey the Luttinger theorem, i.e., 
${\rm Im} \Sigma_{\lambda}(E) \propto (E-\mu)^2$ in the neighborhood of the
Fermi energy.
Due to the cubic symmetry there are two different selfenergies, 
namely, $\Sigma_{t_{2g}}(E)$ and $\Sigma_{e_{g}}(E)$,
corresponding to the three- and two-dimensional representations for 
the $d$-states, respectively.
Another point to be mentioned is that 
$|{\rm Im} \, \Sigma_{t_{2g}}(E)| < |{\rm Im} \, \Sigma_{e_{g}}(E)|$,
because the $e_{g}$ band is narrower than the $t_{2g}$ band, so that 
the ratio $U/w$ is larger for $e_{g}$-states and hence the influence 
of electron interactions is also stronger for $e_{g}$-states.
The states above the Fermi energy are only weakly damped by
electron interactions and are shifted almost rigidly, i.e., 
$\Sigma_{\alpha}(E) \approx {\rm Re} \, \Sigma_{\alpha}(E) \approx {\rm const.}$,
($\alpha=t_{2g},e_{g}$).
The same concerns the low-lying $s$-states at the bottom of the band 
(${\bf k}=\Gamma$) which are influenced by electron-electron interactions 
among the $d$-electrons only indirectly, via a hybridization with the 
$d$-states.
In both cases, the peaks have a Lorentzian shape indicating well-behaved
quasiparticles.
The same holds for the $d$-states close to the Fermi energy both for
${\bf k}=\Gamma$ and ${\bf k}=$X.
On the other hand, the $d$-states well below the Fermi energy, particularly
in the energy region where $|{\rm Im} \, \Sigma_{\alpha}(E)|$ have their maxima 
($E \approx -0.4$~Ry), are strongly influenced by electron interactions.

\epsfxsize=110mm
\vglue 1mm
\epsffile{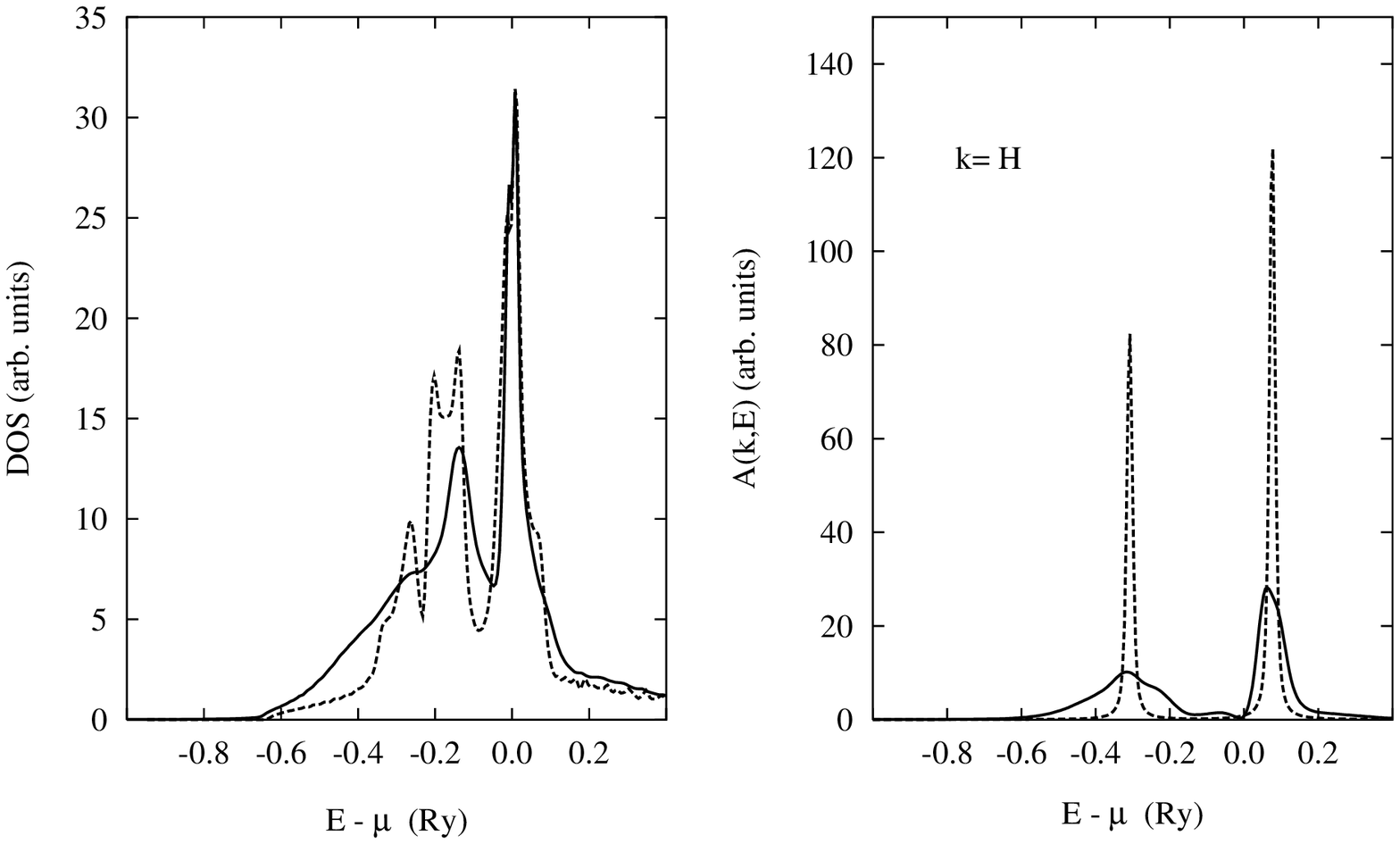}
\vspace*{-15mm}
{\tbf Figure 6.}
{\tenrm  Densites of states (left frame) and spectral densities at 
         ${\bf k}$=H (right frame) for a paramagnetic bcc-Fe within the 
         LDA (dashed lines) and the non-selfconsistent 2nd-order 
         perturbation theory (full lines) for Hubbard parameter $U=0.1$ Ry.}

\hspace{9mm} 

This is illustrated by spectral densities for ${\bf k}$=X at energies
around $E \approx -0.3$~Ry.
The extra spectral weight in the energy region $E \in (-0.6,-0.2)$~Ry 
which is missing within the LDA, gives rise to a satellite structure 
observed in the DOS.
A strongly non-Lorentzian behavior of spectral densities in this energy
region indicates a breakdown of the quasiparticle behavior.
Note a particularly strong influence on the lower $d$-band at ${\bf k}=X$
and a weaker influence on states at ${\bf k}=\Gamma$, where the $d$-states
are closer to the Fermi energy.
We note that along the line $\Gamma$-X, the band-like 
features move with ${\bf k}$-vector, while the extra spectral weight
connected with the satellite structure, remains localized in the BZ.
The change of the sign of ${\rm Re} \Sigma_{\alpha}(E)$ means that the states 
below and above $E \approx -0.4$~Ry are shifted in opposite directions.
The band narrowing can be explained by a negative slope of 
${\rm Re} \, \Sigma_{\alpha}(E)$. 
The results for ${\bf k}=$X agree reasonably well with the 
results of the SOPT calculations \cite{SAS} despite the fact
that in the cited paper the authors studied a ferromagnetic Ni. 
Its exchange splitting is rather small.
We have found that behavior of spectral densities in other approximations 
(TMA, RPA, and GWA) is qualitatively similar and thus need not be 
discussed separately here.

\hspace{9mm} 
We now briefly discuss the results for a bcc Fe to see differences 
due to a different structure, smaller $U/w$ ratio, and a lower electron 
concentration.
The results for the DOS and the spectral density at $\bf k$=(1,0,0)
corresponding to the H-point (see Fig.~4) are plotted in Fig.~6 for the SOPT.
The results are in a reasonable agreement with those of paper \cite{SAS},
although one has to keep in mind that our value of $U$ is slightly
larger and calculations are performed for a paramagnetic rather
than for a ferromagnetic state.
In the LDA, a bcc Fe exhibits well-known bonding and antibonding peaks
with a high value of the DOS at the Fermi energy.
This feature indicating a strong tendency of a paramagnetic Fe 
towards a magnetic instability (Stoner criterion) is preserved in the
SOPT.
On the other hand, the peaks below the Fermi energy are smoothed by 
many-body effects (finite imaginary part of the selfenergy) and due to 
a larger number of holes (partly filled $d$-band) also states above the 
Fermi energy are now influenced.
This is better seen in the spectral density evaluated at $\bf k$=H.
In particular, the LDA peak above the Fermi energy (see Fig.~6) is 
now much more broadened than similar peaks in the case of Ni.
The other peak below the Fermi energy is influenced by electron 
interactions more strongly showing a weak satellite feature around
$E<-0.4$~Ry.

 ~\\
 ~\\
 ~\\
{\bf CONCLUSIONS AND OUTLOOK}\\

\hspace{9mm} 
In summary, we have developed a scheme that allows to determine 
electronic properties of correlated solids which could be described by 
a multiband Hubbard Hamiltonian and which belong to the weak-interaction
case, $U/w<1$. 
Our approach describes the one-electron part of the many-body
Hamiltonian in terms of a TB-LMTO Hamiltonian.
The TB-LMTO is the LDA-based technique which describes reasonably well 
the ground-state properties of solids.
The present formalism is particularly suitable for future extensions 
to random alloys and their surfaces.
This represents a great advantage of the present formulation in
comparison to similar constructions found in the literature.
The many-body part is described by a non-selfconsistent FLEX-type
method, but the Fermi energy is determined selfconsistently.
There are strong arguments \cite{BCW} that such an approach is in
fact more correct than simple selfconsistent treatments using dressed
rather than bare propagators, because within an internally consistent 
theory one has to renormalize simultaneously propagators and vertices.
We are working on such a generalization of single-channel approximations
to include a properly selected subclass of parquet diagrams \cite{VJ}.
In the future, we plan to implement it into the present formalism.
The basic approximation adopted in the present paper is a local
character of the selfenergy.
This limitation which is reasonably justified for transition
metals is also motivated by future applications to alloys and 
their surfaces as it is consistent with the treatment of the 
substitutional disorder using the CPA.
On the other hand, the assumptions concerning the paramagnetic
state, and, more importantly, the simplified form of the electron-electron
interaction reduced here to a single value of the Hubbard parameter $U$
is not essential and can easily be lifted.
We employ retarded quantities evaluated along a line in the complex energy 
plane coupled with an analytic deconvolution to the real axis and combined 
with the use of dispersion relations for the determination of the analytic 
function from its imaginary part.
This is a powerful alternative to the conventional approach based on the 
use of the causal Green functions in which the energy integrals are 
replaced by finite sums over Matsubara energies.
The knowlegde of the electron Green functions then allows one not
only to evaluate one-electron properties, but also a future
extension to transport properties of alloys and multilayers.
 ~\\
 ~\\
 ~\\
{\bf Acknowledgements}\\

\hspace{9mm} 
Financial support for this work was provided by the Grant Agency 
of the Czech Republic (Project 202/98/1290), and
the Grant Agency of the Academy of Sciences of the Czech Republic
(Project A 1010829).
 ~\\
 ~\\
 ~\\
{\bf REFERENCES}\\

\end{document}